\documentclass[
reprint,
superscriptaddress,
showpacs,
preprintnumbers,
showkeys,
amsmath,amssymb,
aps, longbibliography
]{revtex4-1}

\usepackage{graphicx}
\usepackage{dcolumn}
\usepackage{bm}
\usepackage{braket}
\usepackage{color}
\usepackage{float}

\newcommand{\V}[1]{\mathbf{#1}}  
\newcommand{\Up}[1]{{#1}^{\dagger}} 

\begin{document}

\title{Generation of path-encoded Greenberger-Horne-Zeilinger states}

\author{N. Bergamasco}
 \email{nicola.bergamasco01@universitadipavia.it}
 \affiliation{Department of Physics, University of Pavia, Via Bassi 6, I-27100 Pavia, Italy}
\author{M. Menotti}
 \affiliation{Department of Physics, University of Pavia, Via Bassi 6, I-27100 Pavia, Italy}
\author{J. E. Sipe}
 \affiliation{Department of Physics, University of Toronto, 60 St. George Street, Toronto,\\ Ontario M5S 1A7, Canada}
\author{M. Liscidini}
 \affiliation{Department of Physics, University of Pavia, Via Bassi 6, I-27100 Pavia, Italy}

\date{\today}

\begin{abstract}
We study the generation of Greenberger-Horne-Zeilinger (GHZ) states of three path-encoded photons. Inspired by the seminal work of Bouwmeester et al. \cite{Bouwmeester} on polarization-entangled GHZ states, we find a corresponding path representation for the photon states of an optical circuit, identify the elements required for the state generation, and propose a possible implementation of our strategy. Besides the practical advantage of employing an integrated system that can be fabricated with proven lithographic techniques, our example suggests that it is possible to enhance the generation efficiency by using microring resonators.
\end{abstract}


\maketitle

\section{Introduction}
Quantum correlations between subsystems are the focus of many studies on the foundations of quantum mechanics, and the ability to generate states that exhibit these correlations is central to quantum information processing.
While quantum correlations in a bipartite state are generally well-understood \cite{Nielsen}, the analysis of multipartite states is more intricate.
Even for tripartite entangled states, where only three subsystems are involved, one can identify separable states, biseparable states, and two inequivalent classes of tripartite entangled states \cite{Cirac}: Greenberger-Horne-Zeilinger (GHZ) states \cite{GHZ_Paper,GHZ_Book}, and W states \cite{Kiesel}. A state in one class cannot be transformed into one of the other using only local operations and classical communications.

In this communication, we focus on the generation of tripartite GHZ states, the simplest of which can be written as
\begin{equation}\label{GHZ paradigm}
	\ket{GHZ}=\frac{1}{\sqrt{2}}\big(\ket{000}+\ket{111}\big),
\end{equation}
where $\ket{0}$ and $\ket{1}$ are orthogonal states. 
GHZ states were first studied experimentally by Bouwmeester et al. \cite{Bouwmeester}, where the states $\ket{0}$ and $\ket{1}$ identified orthogonal photon polarizations. But other implementations of the orthogonal states are possible, and have been demonstrated in a variety of platforms including trapped ions \cite{Roos} and superconducting circuits \cite{DiCarlo}. GHZ states have been applied in tests of local realism \cite{Pan}, where the use of tripartite states allows for a demonstration of its conflict with quantum mechanics even in a definite measurement, as opposed to such tests using bipartite states which rely on the statistics of a large number of measurements. They have also been used to devise quantum communication protocols, such as multipartite quantum key distribution, with secret keys shared safely among three parties \cite{Jin}; dense coding \cite{Hao}, where the capacity of a transmission channel is increased by using quantum states of light; and entanglement swapping \cite{Xiaolong}.

When photons are used to produce a GHZ state, the entangled degree of freedom is  typically polarization. This choice arises from the fact that polarization can be naturally used as a qubit, and because polarization-entangled photon pairs are now routinely produced by parametric sources \cite{kwiat95}. Moreover, rotation of a polarization-encoded qubit on the Bloch sphere can be easily done by means of linear optical elements such as wave-plates, and routing of the photons can be performed using beam splitters (BSs) and polarizing beam splitters (PBSs).
Yet the use of polarization can be problematic for long distance communication using optical fibers, where polarization can drift during propagation, and for the development of integrated quantum devices, where sophisticated solutions are required to control light polarization on a chip \cite{Matsuda:2012aa}. Thus, the use of other degrees of freedom in photonic implementations of GHZ states is worth investigating.

In this paper we propose a scheme to prepare GHZ states, with the generated photons entangled in the path degree of freedom; the states $\ket{0}$ and $\ket{1}$ here refer to the photon being in different spatial modes \cite{Matthews}, regardless any other degree of freedom. In presenting our strategy we consider, as an example, an integrated optical circuit in which two photon pairs are generated by spontaneous four-wave mixing (SFWM) in a $\chi^{(3)}$ material. While similar schemes could be implemented in different platforms, the approach we suggest allows us to take advantage of the enhancement of the generation rate provided by integrated microresonators, and to drastically reduce the footprint of the source \cite{Azzini:12}.
In principle, it would be possible to design optical schemes that manipulate path-encoded states and subsequently translate and output them in the polarization representation. This has been proposed recently to achieve chip-to-chip quantum communication \cite{Wang16}. However, here we are mainly interested in both the manipulation and output of path-encoded states on optical chips.

\begin{figure}
\includegraphics[width=\columnwidth, keepaspectratio]{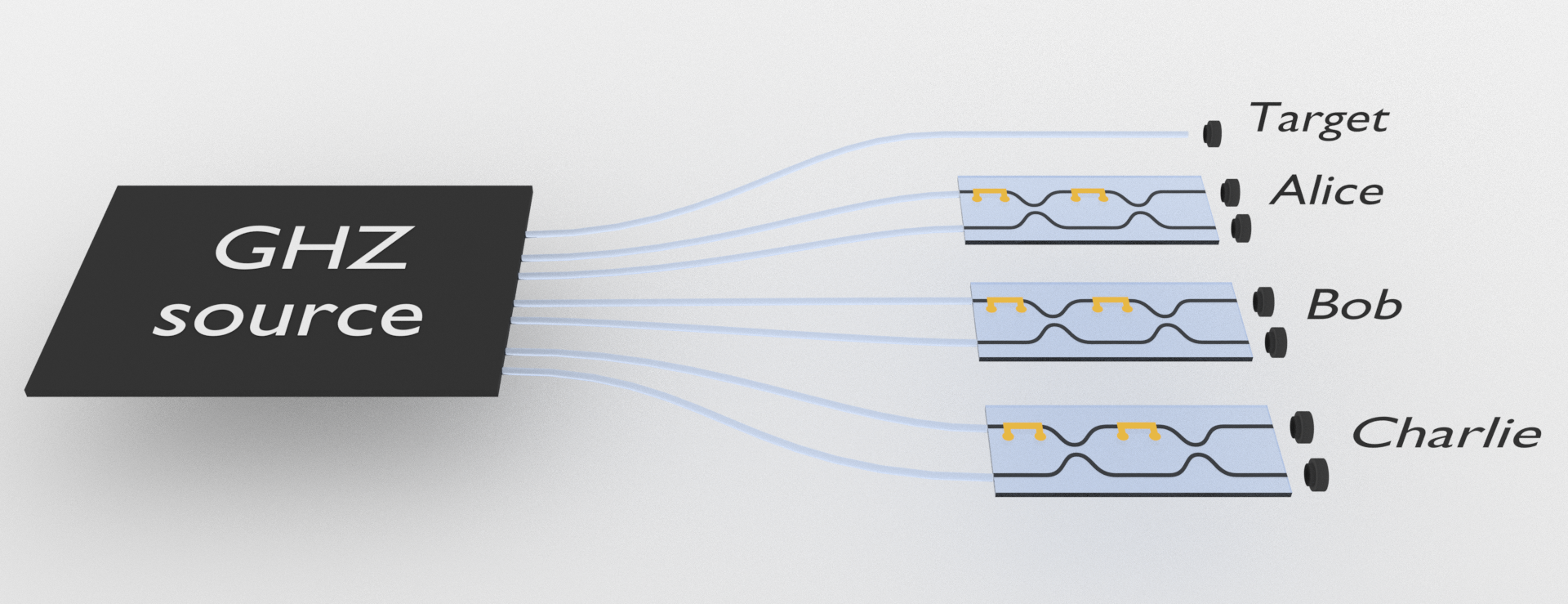}
\caption{\label{ABC} Sketch of the general scheme for the preparation and distribution of path-encoded GHZ states.}
\end{figure}

We envision the situation depicted in Fig. \ref{ABC}: Four photons are generated in an integrated device, of which one is used as a target and three are used as qubits. For each qubit there are two paths, each path associated with a basis state. The three photons are routed to three independent parties (Alice, Bob, and Charlie), which can manipulate them, where the rotation of the qubit on the Bloch sphere is performed by means of a Mach-Zehnder interferometer and two phase shifts \cite{Silverstone15}.

The work is organized as follows: in section 2 we establish a correspondence between some relevant optical elements used in polarization optics and the integrated 
counterparts in the path-encoding framework. In section 3 we present the integrated approach for generating the GHZ state, we discuss how to post-select the desired state, and we estimate the generation rate. Finally, in section 4 we draw our conclusions.

\section{From polarization- to path-encoded qubits}
Bulk sources used to generate quantum correlated photons typically rely on spontaneous parametric down-conversion (SPDC) in nonlinear crystals, e.g. $\beta$-barium borate (BBO) \cite{kwiat95}, or on SFWM, e.g. in optical fibers \cite{Smith:2009}. The former is a second-order nonlinear process that can be pictured as the spontaneous fission of a pump photon into two daughter photons of lower energy, while the latter is a third-order nonlinear process that can be regarded as the elastic scattering of two pump photons to yield a new photon pair. These two processes can also be used in photonic integrated circuits (PICs) \cite{Azzini:12,Ducci:2013}. In this context, SFWM is particularly useful, for the circuit can be easily fabricated in silicon, with recent implementations employing silicon nitride \cite{Moss}. These materials possess a relatively strong third-order nonlinear susceptibility that favours SFWM, and also provide strong field confinement thanks to the large index contrast with silicon dioxide, which is usually used as the low-index cladding material in the fabrication of ridge waveguides and resonators. In principle, the polarization of the generated photons can be used to implement a qubit either in a bulk or integrated source. Yet in PICs this is particularly challenging, and thus alternative solutions are desirable \cite{Menotti:2016}.

\begin{figure}
\includegraphics[scale=0.35,keepaspectratio]{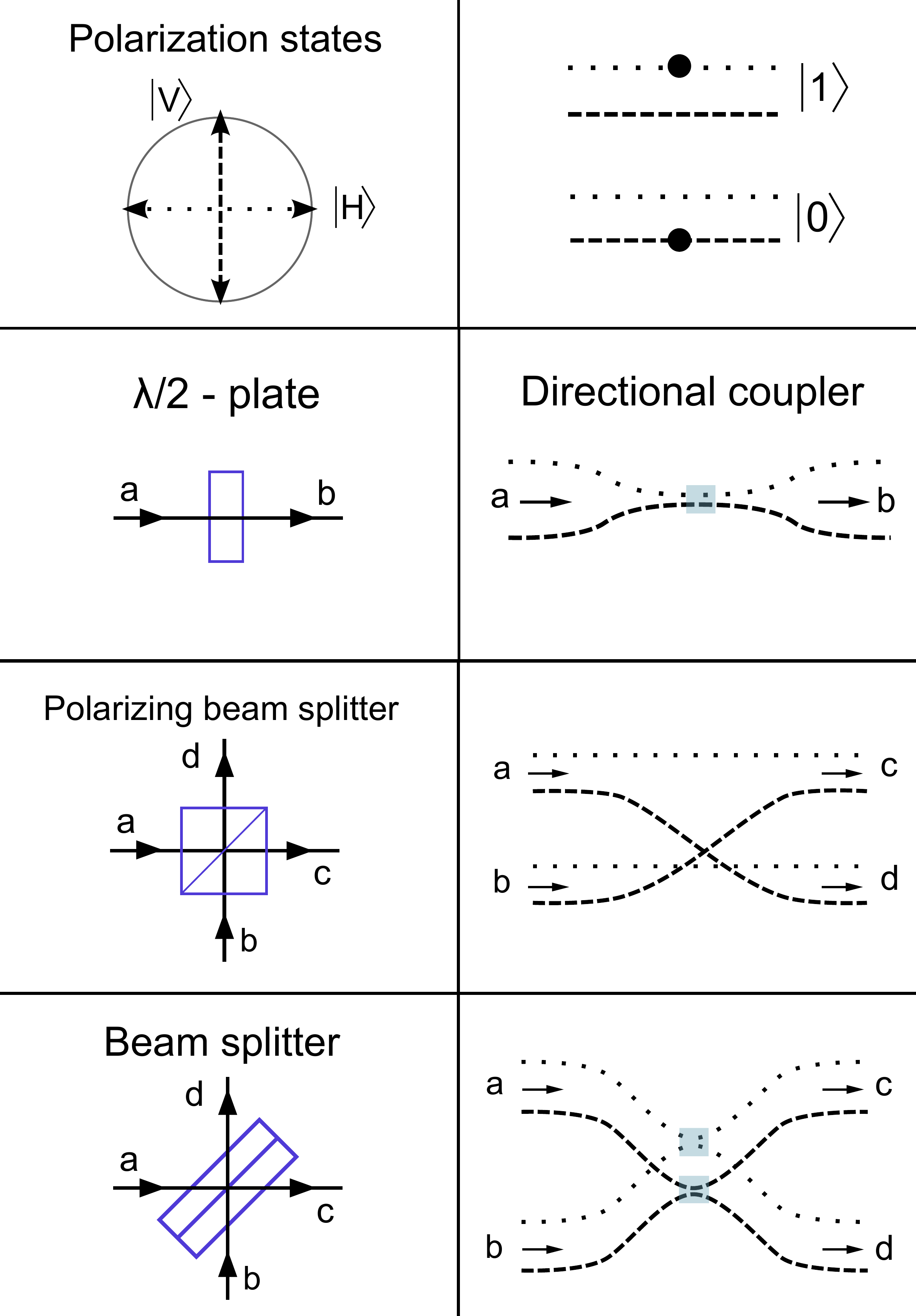}
\caption{\label{Analogies}Analogies between optical elements employed in bulk optics for schemes involving polarization-entangled states (on the left) and the corresponding integrated optical elements for the scheme introduced here involving path-encoded states (on the right). Dotted and dashed lines indicate waveguides associated with $\ket{1}$ and $\ket{0}$, respectively. The shaded boxes mark the coupling points between waveguides.}
\end{figure}

In this section we investigate the possibility of using the path degree of freedom of photons for qubit encoding. To this end, we propose employing two waveguides, or \emph{paths}, for each photon route in a PIC. We assign the state $\ket{1}$ or $\ket{0}$ to a photon when it travels in one waveguide or the other, which we graphically depict as dotted and dashed, respectively, in Fig. \ref{Analogies}. This convention is kept consistent throughout the whole circuit.
 
In Fig. \ref{Analogies} we show that there is a full correspondence between bulk optical elements used to manipulate polarization states and integrated optical elements necessary to manipulate path-encoded states.

The rotation of polarization states is performed in bulk optics by using a $\lambda/2$ plate, while the corresponding evolution of path states is effected with a 50:50 directional coupler (DC) connecting the two waveguides associated with the $\ket{1}$ and $\ket{0}$ states. 
Photons in a bulk optical circuit can be routed depending on their polarization using a PBS; the same can be done for the path-encoded states  by properly connecting the waveguides of the input ports to the waveguides of the output ports (see Fig. \ref{Analogies}). Finally, photons in a bulk optical circuit can be spatially separated regardless their polarization state using a BS, and the corresponding operation on path-encoded states is performed in integrated optics using two 50:50 DCs.

Two remarks regarding path states and their manipulation are necessary: First, we note that the generation of a meaningful path-encoded state for a photon pair requires a source more complicated than a single-bus-waveguide ring resonator \cite{Mataloni04, Silverstone14, Preble15}. 
Second, some of the integrated optical elements used to manipulate path states (see Fig. \ref{Analogies}) display a waveguide crossing that seems problematic in a planar geometry, which is usually the choice for PICs. However, we will see that proper sources can be designed, and a waveguide rearrangement can avoid the problematic waveguide crossing.

\section{State generation and manipulation}
Here we discuss the generation of path-encoded GHZ states and present an integrated circuit based on the fundamental building blocks introduced in the previous section.
Considering a generic parametric source, in the approximation of undepleted pump pulses described classically, the state of the generated photons is of the form \cite{braunstein2005}
\begin{multline}
\ket{\psi} = e^{\beta C^\dagger_{II}-H.c.}\ket{\text{vac}}\\
	      = \left(1+\mathcal{O}(|\beta|^2)\right)\ket{\text{vac}} + \beta\Up{C}_{II}\ket{\text{vac}} + \frac{1}{2}\left[\beta\Up{C}_{II}\right]^2\ket{\text{vac}}+ \ldots \\
	      \equiv \left(1+\mathcal{O}(|\beta|^2)\right)\ket{\text{vac}} + \beta\ket{\text{II}} + \frac{1}{2}\left[\beta\Up{C}_{II}\right]^2\ket{\text{vac}} + \ldots,
\label{eq:sfwmstate}
\end{multline}
where $\ket{\text{vac}}$ is the vacuum state, $\Up{C}_{II}$ is the photon pair creation operator, $\left|\beta\right|^2$ is the pair 
generation probability per pulse when that number is very small, and $\ket{\text{II}}$ is the normalized two-photon state. In the limit of interest where $\left|\beta\right|^2\ll1$, we can truncate the expansion \eqref{eq:sfwmstate} at the quadratic term in $\beta$ , which corresponds to the generation of two pairs.
The properties of the four-photon state contribution to \eqref{eq:sfwmstate}, resulting from the generation of two photon pairs, are directly related to the those of the creation operator $\Up{C}_{II}$. Hence, once this has been calculated, the output state of two or more pairs can be obtained immediately.
For this reason, we begin with a discussion of the generation of a single photon pair.

The structure we propose can be divided in two parts: a nonlinear integrated source, which generates a path-encoded initial state, and a linear optical circuit to manipulate it. The full calculation of the output state is reported in the Appendix.
\begin{figure}[H]
\includegraphics[scale=0.4, keepaspectratio]{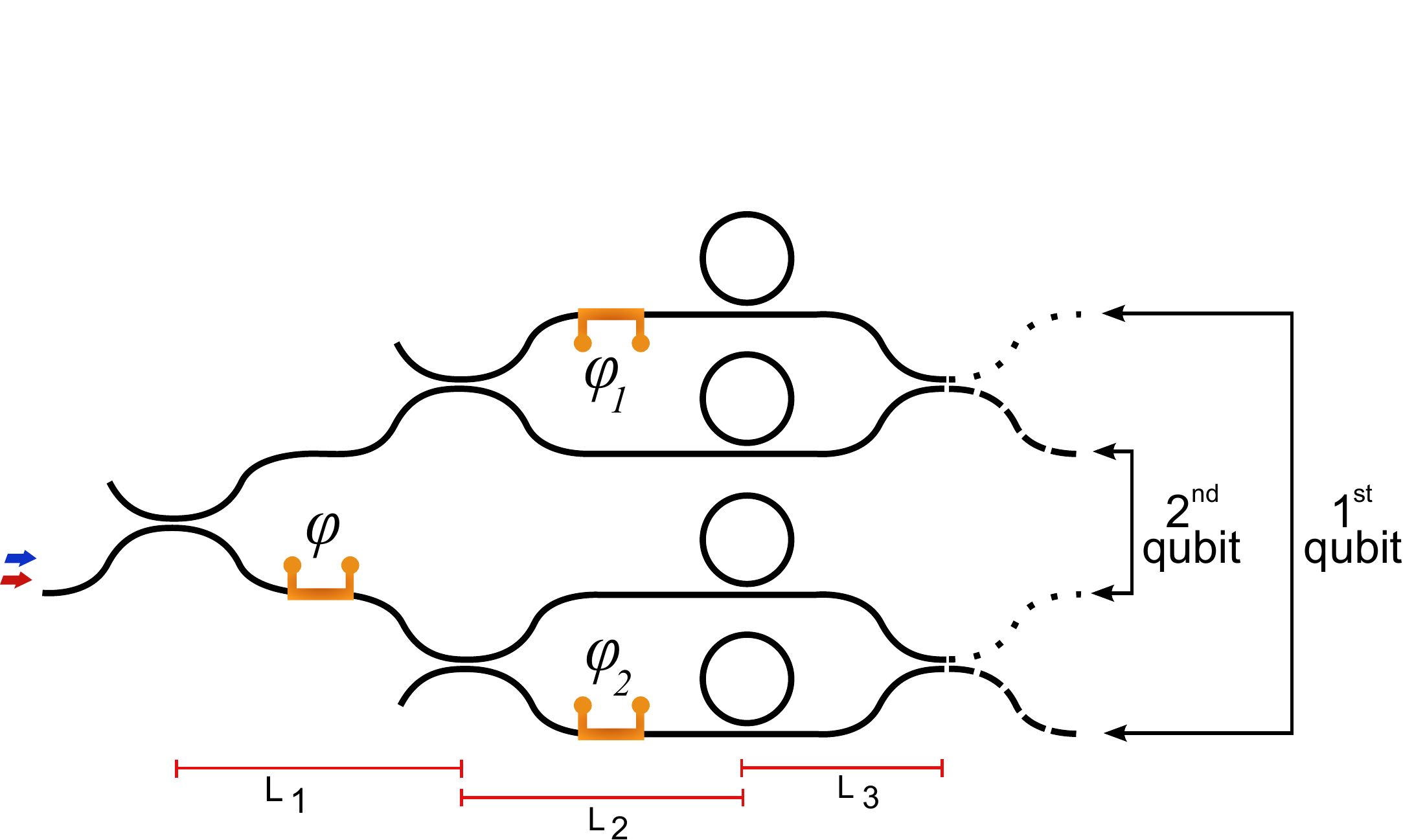}
\caption{\label{source}Sketch of the nonlinear integrated source of path-entangled states \eqref{eq:pathent}. Waveguides associated with single qubits are grouped together, phase shifters and relevant lengths are also shown.}
\end{figure}
The nonlinear integrated source (see Fig. \ref{source}) consists of four identical ring resonators arranged in two blocks, each of which is a Mach-Zehnder interferometer unbalanced by a phase $\phi_i$, with one ring resonator per arm. The two blocks are coherently pumped using a 50:50 directional coupler, which splits the pump amplitude into two waveguides, with $\phi$ being the pump phase difference between the two blocks. Although this is not strictly necessary, here we consider degenerate SFWM \cite{Fang:13}, for which we require a dual-pump scheme, where the 50:50 split ratio can be guaranteed by choosing an appropriate length of the directional coupler \cite{Huang:94}. Since the field enhancement inside the rings is much larger than that in the waveguides, we assume that the generation of photons occurs only in the resonators.

%Finally, although not strictly necessary, we assume to work with degenerate photons, assuming a dual pump configuration \cite{Fang:13}.

It should be noticed that although the use of four identical microring resonators might pose some constraints, the fabrication technique for multiple integrated elements on SOI platforms has constantly improved in recent years, up to the realization of several hundred coupled microrings \cite{Mookherjea}. Moreover, it is possible to tune each resonator almost independently via heaters: this enables the control of the position of its resonances with great precision \cite{Cunningham:10}. If one considers silicon ring resonators, the large nonlinearity ($\gamma\approx 200\ W^{-1}m^{-1}$) guarantees high generation efficiencies with mW pump powers and $Q\approx10000$ \cite{Azzini:12}, which relaxes the constraints on the ring tunability. Finally, the two blocks in Fig. \ref{source} have already been used for the generation of deterministically split photons by the reverse HOM effect, yielding high-visibility quantum interference \cite{Silverstone15}.
Indeed, when $\phi_i=\pi/2[2\pi]$ one observes deterministic splitting of the photon pair exiting the MZI \cite{Silverstone14}. But when the two blocks are pumped with a relative phase shift $\phi=\pi$ (or odd multiple), the two-photon state generated by the source is the Bell state (see the Appendix)
\begin{equation}
\ket{\Psi^-} = \frac{1}{\sqrt{2}}\left(\ket{1}\ket{0} - \ket{0}\ket{1}\right),
\label{eq:pathent}
\end{equation}
where we use the first and fourth waveguide for the first path-encoded qubit and we use the second and the third waveguide for the second path-encoded qubit as depicted in Fig. \ref{source}. This situation is analogous to that considered by Bouwmeester et al. \cite{Bouwmeester}, where the nonlinear crystal generates photon pairs in the corresponding polarization-encoded entangled state.

We now consider the simultaneous generation of two pairs of photons, described by the effect of $(C_{II}^\dagger)^2$ on the vacuum state. This leads to the four-photon state
\begin{multline}
\ket{\mathrm{IV}} = -\frac{1}{2\sqrt{3}}\int dk'_1dk'_2 dk_1dk_2
\phi_{\text{ring}}(k_1,k_2)\phi_{\text{ring}}(k'_1,k'_2)\\
\times e^{i(\psi(k_1,k_2) +\psi'(k_1,k_2))}(\Up{b}_{k_1,1}\Up{b}_{k_2,2}
- \Up{b}_{k_1,3}\Up{b}_{k_2,4})\\
\times (\Up{b}_{k'_1,1}\Up{b}_{k'_2,2}
- \Up{b}_{k'_1,3}\Up{b}_{k'_2,4})\ket{\text{vac}},
\label{eq:fourphostate_text}
\end{multline}
where $\phi_{\text{ring}}(k_1,k_2)$ is the biphoton wave function of a pair generated in a single ring, $\psi(k_1,k_2)$ and $\psi^{\prime}(k_1,k_2)$ are phase factors associated with propagation in the channel (which can be assumed constant) defined in \eqref{eq:psiwave}, and $b^{\dagger}_{k_i,j}$ is the operator associated with the creation of a photon having wavevector $k_i$ and exiting  the structure in Fig. \ref{source} from the channel $j$. The state $\ket{\text{IV}}$ is normalized under the assumption that the biphoton wave function $\phi_{\text{ring}}(k_1,k_2)$ is separable (see below).

\begin{figure}[H]
\includegraphics[scale=0.3, keepaspectratio]{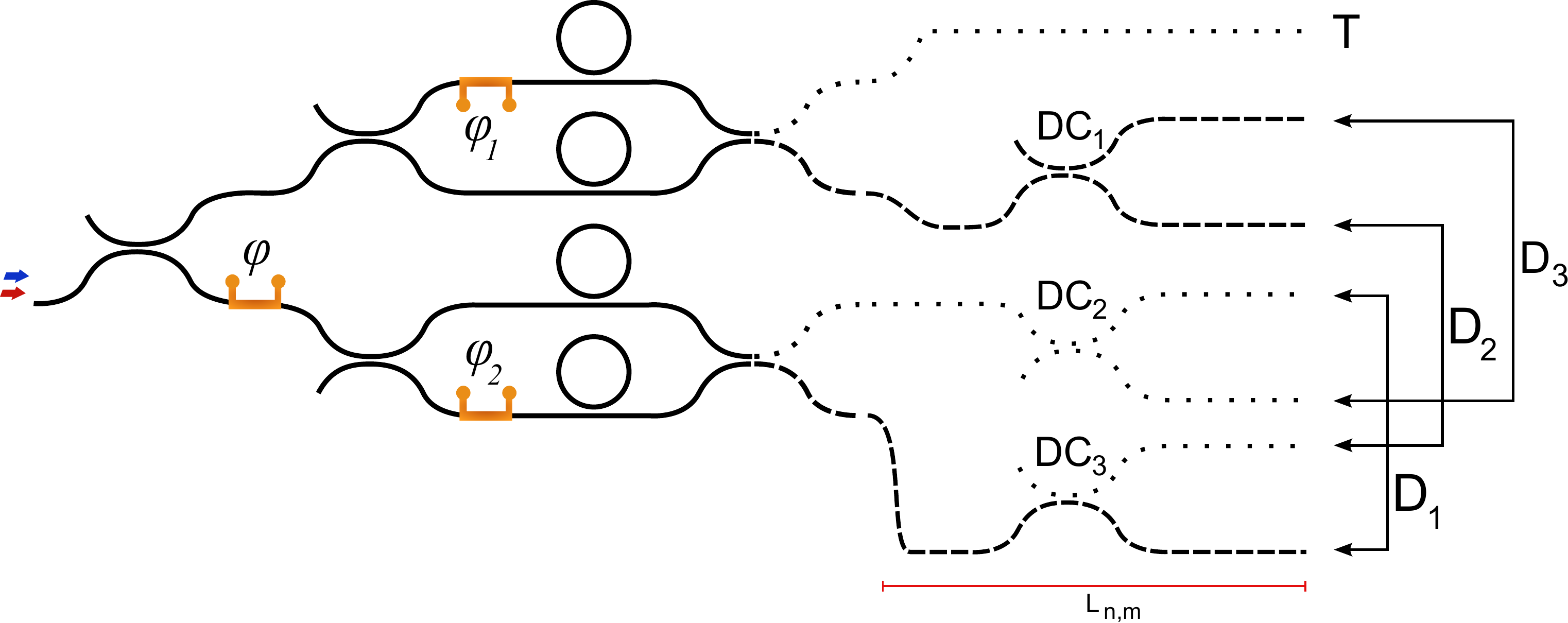}
\caption{\label{schemfull}Sketch of the complete integrated circuit for the generation of path-encoded GHZ states. We can identify 
a schematic representation of the source of path-encoded entangled states in the form \eqref{eq:pathent}, and a realistic implementation of the full circuit obtained by rearranging some of the channels and the detectors. Waveguides associated with single qubits are grouped together.}
\end{figure}

We now turn to the manipulation of the state $\ket{\mathrm{IV}}$, which is done following the recipe Bouwmeester et al. \cite{Bouwmeester} used for polarization-encoded entangled states, but implemented for path-encoded entangled states using the correspondence between polarization bulk elements and the path integrated components illustrated in Fig. \ref{Analogies}. Note that we have avoided the waveguide overlapping in the integrated analogue of a beam splitter (see Fig. \ref{Analogies}) by a rearrangement of the circuit waveguides as shown in Fig. \ref{schemfull}. 

In strict analogy with Bouwmeester et al. \cite{Bouwmeester}, post-selecting on a three-fold coincidence in detectors $D_1$, $D_2$, and $D_3$ in Fig. \ref{schemfull}, conditioned on the detection of a photon in the target detector $T$, identifies that a GHZ state was generated. Care must be taken to ensure that the generated GHZ state is pure. As in the generation of pairs of photons for heralded photon applications, this requires that the function $\phi_{\text{ring}}(k_1,k_2)$ is separable. To this end we observe that nearly-uncorrelated photons can be obtained by adjusting the duration of the pump pulse, which has to be comparable or shorter than the dwelling time of the  photon inside the ring \cite{Helt10,onodera16}. For complete separability, a more careful design of the ring is required \cite{Gentry:16,Vernon:2017}. A more detailed discussion of the effect of the spectral properties of the BWF goes beyond the scope of this work, but we plan to examine this issue in a future communication.

Following earlier work \cite{Liscidini12}, the state \eqref{eq:fourphostate_text} can be written in terms of the creation operators corresponding to the asymptotic-out field of the structure of Fig. \ref{schemfull}. The relation between the asymptotic-in and -out field operators is reported in \eqref{eq:evo} of the Appendix. This allows us to rewrite the complete output state as:
\begin{align}
\ket{\psi} &= \left(1+\mathcal{O}(|\beta|^2)\right)\ket{\text{vac}} \nonumber \\
	      &+ \beta\ket{\text{II}} + \frac{\sqrt{3}}{2}\beta^2\left[\ket{\Phi} -\frac{1}{2\sqrt{3}}\ket{\psi_{\text{GHZ}}}\right],
\label{eq:outstate}
\end{align}
where $\ket{\Phi}$ includes other contributions that are second order in $\beta$ but would not lead to a four-fold coincidence event, while 
\begin{multline}
\ket{\psi_{\text{GHZ}}} = \int dk_1dk_2dk'_1dk'_2\phi_{\text{ring}}(k_1,k_2)
\phi_{\text{ring}}(k'_1,k'_2)\\
\times e^{i\Gamma}\ket{\text{T}}\ket{\text{GHZ}},
\label{eq:ghz5}
\end{multline}
with $\Gamma$ a phase factor and
\begin{align}
\ket{\text{GHZ}}  &= \frac{1}{\sqrt{2}}\left[\Up{b}_{k_1,D_{1,1}} \nonumber
\Up{b}_{k_2,D_{2,1}}\Up{b}_{k'_2,D_{3,0}} \right.\nonumber \\
&+ \left. e^{i\Theta(k_2,k'_1,k'_2)}\Up{b}_{k_2,D_{1,0}}\Up{b}_{k'_2,D_{2,0}}\Up{b}_{k_1,D_{3,1}}\right]\ket{\text{vac}}\nonumber \\
 &= \frac{1}{\sqrt{2}}\left[\ket{110} + e^{i\Theta}\ket{001}\right], \qquad\qquad\qquad \nonumber\\
 \ket{T}&=\Up{b}_{k'_1,T}\ket{\mathrm{vac}}.
\label{eq:GHZ}
\end{align}
Here $\Theta(k_2,k'_1,k'_2)$ is a relative phase between the two GHZ state components and depends on the relative positions of the detectors (see \eqref{eq:Theta}), which cannot be longer than the coherence length of the photons. Such a coherence length can always be increased by filtering, although for typical resonance widths achievable at telecom wavelengths in silicon and silicon nitride resonators it already ranges from centimetres to meters \cite{Moss, Grassani15, Preble15}.

As expected, any four-fold coincidence event results in a GHZ state, where the probability of such an event is $\left|\beta^2/4\right|^2$, when propagation losses are neglected. The magnitude  of $\beta$ depends on the pump power, the ring radius and the quality factor of the resonators, and it can vary depending on the device under consideration. Yet values of $\left|\beta\right|^2 \approx 0.1$ have been demonstrated in PICs \cite{Silverstone15}, and assuming MHz pump repetition rates, this would allow for the preparation of path-encoded GHZ states with kHz generation rates with mW pump powers and quality factors of the order of $10^4$. Although our theoretical estimate does not account for any loss, device imperfection, and detector efficiencies, we still expect a large improvement on the generation rate with respect to the present results reported in the literature.

\section{Conclusions}
For the generation and processing of quantum correlated photons, we have shown that there is a one-to-one correspondence between components operating in a path encoding scheme and bulk optical elements operating in a polarization encoding scheme. Exploiting this result, we proposed and studied the generation of path-encoded tripartite GHZ states. Although the generation of the desired state is revealed only in post-selection, therefore destroying the quantum state, many protocols involving GHZ states are based on this condition \cite{Jin,Hao,Xiaolong}.
Our approach is suitable for the generation of multipartite states in quantum photonic integrated devices, as it overcomes the difficulties related to the use of the polarization degree of freedom. To demonstrate this, we designed and studied an integrated structure relying on the generation of photon pairs by SFWM in ring resonators, showing the potential of this approach in terms of source footprint and brightness.

\section*{Acknowledgments}
We are grateful to Mario Arnolfo Ciampini for the critical reading and the fruitful discussions on the manuscript.

\appendix
\section{}
Referring to the schematic representation in Fig. \ref{source}, the photon pair creation operator $C_{II}^\dagger$ can be expressed very generally as
\begin{equation}
C^\dagger_{II} = \frac{1}{\sqrt{2}}\sum_{p,q}\int dk_1dk_2\phi_{p,q}\left(k_1,k_2\right)b^\dagger_{k_1,p}b^\dagger_{k_2,q},
\end{equation}
where $p$ and $q$ run over all the output channels, $\phi_{p,q}$ is the amplitude of the biphoton wave function (BWF) that is associated with the photon pair exiting from channels $p$ and $q$ and
\begin{equation}\label{eq:normalcondition}
\sum_{p,q} \int\ dk_1dk_2 \left| \phi_{p,q}(k_1,k_2)\right|^2=1.
\end{equation}

The particular arrangement of the ring resonators in our source design allows for only a restricted number of combinations in $(p,q)$:
\begin{multline}
(p,q)\in\Omega = \left\{(1,1);(2,1);(1,2);(2,2);\right. \\ \left.(3,3);(4,3);(3,4);(4,4)\right\}.
\end{multline}

To lowest order in the pump intensities, $\phi_{p,q}(k_1,k_2)$ can be written as \cite{Yang, Helt10, Liscidini12}
\begin{multline}\label{eq:phipq}
\phi_{p,q}\left(k_1,k_2\right) =\frac{2\sqrt{2}\pi\alpha^2i}{\beta\hbar}\int dk
\phi_P(k)\phi_P(k_1+k_2-k)\\
\times S_{p,q}\left(k_1+k_2-k,k,k_1,k_2\right),
\end{multline}
where the coupling term $S_{p,q}$ is related to the superposition of the asymptotic-in fields in the structure by
\begin{multline}
S_{p,q}\left(k_1+k_2-k,k,k_1,k_2\right) =\\
= \frac{3}{2\epsilon_0}
\sqrt{\frac{(\hbar\omega_{k_1+k_2-k})(\hbar\omega_{k})(\hbar\omega_{k_1,p})(\hbar\omega_{k_2,q})}{16}}\\
\times\int d\V{r}\Gamma^{ijkl}\left(\V{r}\right)D^{i,\text{asy-in}}_{k_1+k_2-k}(\V{r})D^{j,\text{asy-in}}_{k}(\V{r})\\
\times D^{k,\text{asy-in}}_{k_1,p}(\V{r})D^{l,\text{asy-in}}_{k_2,q}(\V{r}),
\label{Spq}
\end{multline}

\noindent where $\Gamma^{ijkl}\left(\V{r}\right)$ is related to the third-order nonlinear susceptibility tensor \cite{Helt10}.
Working out the explicit form of each term in equation \eqref{Spq} with respect to the scheme in Fig. \ref{source}, we find that
\begin{multline}\label{Spq_sum}
S_{p,q}\left(k_1+k_2-k,k,k_1,k_2\right) =\\
= \frac{3}{2\epsilon_0}\sqrt{\frac{(\hbar\omega_{k_1+k_2-k})(\hbar\omega_{k})(\hbar\omega_{k_1,p})(\hbar\omega_{k_2,q})}{16}}\\
\times\sum_{n\in[1,4]}A_n(k_1+k_2-k)A_n(k)B_{n,p}(k_1)B_{n,q}(k_2)\\
\times\bar{\jmath}(k_1+k_2-k,k,k_1,k_2),
\end{multline}

\noindent where $\bar{\jmath}(k_1+k_2-k,k,k_1,k_2)$ is the overlap integral of the asymptotic-in fields $\tilde{D}_k(\mathbf{r})$ in a single ring
\begin{multline}
\bar{\jmath}\left(k_1+k_2-k,k,k_1,k_2\right)=\int_{1^{st}\text{ ring}} d\V{r}\Gamma^{ijkl}(\V{r})\\
\times \tilde{D}^{i}_{k_1+k_2-k}(\V{r})\tilde{D}^{j}_{k}(\V{r})\tilde{D}^{k}_{k_1}(\V{r})\tilde{D}^{l}_{k_2}(\V{r}),
\label{jbar}
\end{multline}
and the coefficients $A_n$ and $B_{n,p(q)}$ in equation \eqref{Spq_sum} are given by
\begin{align}
A_1(k) &= \left(it\right)^2 e^{i\phi_1}e^{ik(L_1+L_2)},\notag\\
A_2(k) &= itre^{ik(L_1+L_2)},\notag\\
A_3(k) &= r^2 e^{i\phi}e^{ik(L_1+L_2)},\notag\\
A_4(k) &= itr\,e^{i(\phi+\phi_2)}e^{ik(L_1+L_2)}
\label{eq:Acoeff}
\end{align}
and
\begin{align}
B_{1,1}(k) &= r\,e^{-ikL_3},\notag\\
B_{2,2}(k) &= r\,e^{-ikL_3},\notag\\
B_{2,1}(k) &= it\,e^{-ikL_3},\notag\\
B_{1,2}(k) &= it\,e^{-ikL_3},
\label{eq:Bcoeff}
\end{align}
where $L_1$, $L_2$, and $L_3$ are the distances between coupling points, and $\phi$, $\phi_1$, and $\phi_2$ are phase delays (see Fig. \ref{source}).
Summing all the contributions in equation \eqref{Spq_sum} we find that, when all the directional couplers have a 50:50 split ratio and the phase $\phi_1=\phi_2=\frac{\pi}{2}$, the only nonvanishing terms in \eqref{eq:phipq} are
\begin{align}
\phi_{1,2}(k_1,k_2) &= \phi_{2,1}(k_1,k_2)\notag \\
&= \frac{-i}{4}e^{i\psi\left(k_1,k_2\right)}\frac{\beta_{ring}}{\beta}\phi_{ring}(k_1,k_2)
\label{eq:phi12}
\end{align}
and
\begin{align}\label{eq:phi34}
\phi_{3,4}(k_1,k_2) &= \phi_{4,3}(k_1,k_2) \notag \\
&= \frac{i}{4}e^{i\psi\left(k_1,k_2\right)}e^{2i\phi}\frac{\beta_{ring}}{\beta}\phi_{ring}(k_1,k_2),
\end{align}
where $|\beta_{ring}|^2$ is the probability of generating a pair in a single ring resonator, with a BWF \cite{Helt10}
\begin{multline}
\phi_{ring}(k_1,k_2) = \frac{2\sqrt{2}\pi\alpha^2i}{\beta_{\text{ring}}\hbar}\int dk
\phi_P(k)\phi_P(k_1+k_2-k)\\
\times \frac{3}{4\epsilon_0}\sqrt{\frac{(\hbar\omega_{k_1+k_2-k})(\hbar\omega_{k})(\hbar\omega_{k_1})(\hbar\omega_{k_2})}{16}}\\
\times \bar{\jmath}(k_1+k_2-k,k,k_1,k_2)
\end{multline}
which is normalized according to
\begin{equation}\label{eq:phiringnorm}
\int\ dk_1dk_2\left|\phi_{ring}(k_1,k_2)\right|^2=1,
\end{equation}
and where we defined $\psi(k_1,k_2)$ as
\begin{equation}
\psi(k_1,k_2) = 2(k_1+k_2)(L_1+L_2-L_3).
\label{eq:psiwave}
\end{equation}

Now we can finally reconstruct the complete output state generated by the source.
Considering the limit of low generation probability, the ket \eqref{eq:sfwmstate} takes the form
\begin{align}
\ket{\psi} &\approx \left(1+\mathcal{O}(|\beta|^2)\right)\ket{\text{vac}}+\beta C_{II}^\dagger\ket{\text{vac}}\notag\\
&+\frac{1}{2}\left(\beta C_{II}^\dagger\right)^2\ket{\text{vac}}+\cdots \notag\\
			&\equiv   \left(1+\mathcal{O}(|\beta|^2)\right)\ket{vac}+\beta \ket{II}+\frac{\sqrt{3}}{2}\beta^2\ket{IV}+\cdots
\label{eq:expansion}
\end{align}
where the factor $\sqrt{3}$ comes form the normalization of the state $\ket{\text{IV}}$, and the normalized two-photon state is
\begin{multline}
\ket{II} = -\frac{i}{4\sqrt{2}}\int dk_1dk_2\frac{\beta_{\text{ring}}}{\beta}\phi_{\text{ring}}\left(k_1,k_2\right)\\
\times e^{i\psi\left(k_1,k_2\right)}\left\{\Up{b}_{k_1,1}\Up{b}_{k_2,2} + \Up{b}_{k_1,2}\Up{b}_{k_2,1}\right. \\
\left.- e^{2i\phi}\left(\Up{b}_{k_1,3}\Up{b}_{k_2,4}+\Up{b}_{k_1,4}\Up{b}_{k_2,3}\right)\right\}\ket{\text{vac}}.
\label{eq:twophostate3}
\end{multline}
When $\phi=\pi$, \eqref{eq:twophostate3} becomes 
\begin{multline}
\ket{II}  = -\frac{i}{2\sqrt{2}}\int dk_1dk_2\frac{\beta_{\text{ring}}}{\beta}\phi_{\text{ring}}(k_1,k_2)
 e^{i\psi\left(k_1,k_2\right)}\\
\times\left\{\Up{b}_{k_1,1}\Up{b}_{k_2,2}
- \Up{b}_{k_1,3}\Up{b}_{k_2,4}\right\}\ket{\text{vac}},
\label{eq:twophostate4}
\end{multline}
which is equivalent to the Bell state \eqref{eq:pathent} in the path-encoding notation.
It should be noticed that, from the normalization condition \eqref{eq:normalcondition} and equations \eqref{eq:phi12} and \eqref{eq:phi34}, we have
\begin{equation}
4\times\int\ dk_1dk_2 \frac{1}{16}\left|\frac{\beta_{\text{ring}}}{\beta}\right|^2\left| \phi_{\text{ring}}(k_1,k_2)\right|^2=1
\end{equation}
that, together with the normalization condition on the BWF \eqref{eq:phiringnorm}, gives
\begin{equation}
\left|\beta\right|^2=\frac{\left|\beta_{ring}\right|^2}{4}.
\label{eq:efficiency}
\end{equation}

In this context we are interested in the simultaneous generation of two photon pairs, and thus we focus on the next term in the expansion \eqref{eq:expansion}, which involves the four-photon state $\ket{\mathrm{IV}}$; using Eq. \eqref{eq:efficiency}, this leads to Eq. \eqref{eq:fourphostate_text}.

Referring to Fig. \ref{schemfull} and following the notation \cite{Heebner04} for directional couplers, we can express the photon creation operators in \eqref{eq:fourphostate_text} in terms of the photon creation operators in each detector channel $D_{n,m}$ as

\begin{align}\label{eq:evo}
\Up{b}_{k_1,1} &= e^{-ik_1L_T}\Up{b}_{k_1,T} \\
\Up{b}_{k_2,2} &= -it_1e^{-ik_2L_{3,0}}\Up{b}_{k_2,D_{3,0}}+ r_1e^{-ik_2L_{2,0}}\Up{b}_{k_2,D_{2,0}} \notag \\
\Up{b}_{k_1,3} &= -it_2e^{-ik_1L_{3,1}}\Up{b}_{k_1,D_{3,1}}+ r_2e^{-ik_1L_{1,1}}\Up{b}_{k_1,D_{1,1}} \notag \\
\Up{b}_{k_2,4} &= -it_3e^{-ik_2L_{2,1}}\Up{b}_{k_2,D_{2,1}}+ r_3e^{ik_2L_{1,0}}\Up{b}_{k_2,D_{1,0}},\notag 
\end{align}
where $L_{n,m}$ is the distance between the appropriate output directional coupler in the source and the detector $D_{n,m}$, and $L_T$ is the length between the upper directional coupler in the source and the target detector $T$.
Using \eqref{eq:evo} in \eqref{eq:fourphostate_text}, and referring to the output state expansion \eqref{eq:expansion} we find that the state at the detectors is \eqref{eq:outstate}-\eqref{eq:GHZ}, with the relative phase between the terms in the GHZ given by
\begin{align}\label{eq:Theta}
\Theta &= k_1(L_{1,1}-L_{3,1} )+k_2(L_{2,1}-L_{1,0} ) \notag\\
	   &+ k'_2(L_{3,0}-L_{2,0})+\frac{\pi}{2}.
\end{align}

\bibliography{text}

\end{document}